\documentclass[12pt,draftcls,journal,leterpaper,twosides,onecolumn]{IEEEtran}
\usepackage[english]{babel}
\usepackage{amsmath,amssymb,amscd,latexsym,dsfont}
\usepackage{float,color,graphicx,subfigure}
\usepackage{multirow,multicol}
\usepackage{comment,cite}
\usepackage{enumerate}
\usepackage{stfloats}
\usepackage{psfrag}
\usepackage{cite}

\IEEEoverridecommandlockouts

\title{Variable-rate Retransmissions for Incremental Redundancy Hybrid ARQ}
\author{%
\authorblockN{Leszek Szczecinski, Ciro Correa\authorrefmark{3}, and Luciano Ahumada\authorrefmark{3}\\}
\authorblockA{INRS-EMT, Montreal, Canada\\}
\authorblockA{\authorrefmark{3}Escuela de Ingenier\'{\i}a Inform\'{a}tica, Universidad Diego Portales, Santiago, Chile\\}

\emph{leszek@emt.inrs.ca, ciro.correa@mail.udp.cl, luciano.ahumada@mail.udp.cl}
\thanks{This work was supported by the 7th framework program of European Community FP7/2007-2013 under the grant \#236068, and by Fondecyt under grant  \#1095139, and Anillo ACT-53/2010. When this work was submitted for publication, L.~Szczecinski was on sabbatical leave with CNRS, Laboratory of Signals and Systems, Gif-sur-Yvette, France. The results were presented in part at IEEE Global Communication Conference, 6-10 Dec. 2010, Miami, USA.}
}%

\newcommand{\tr}[1]{\mathrm{#1}}
\newcommand{\mb}[1]{\mathbf{#1}}
\newcommand{\ov}[1]{\overline{#1}}

\newcommand{\mc}[1]{\mathcal{#1}}

\newcommand{\bs}[1]{\boldsymbol{#1}}
\newcommand{\set}[1]{\{#1\}}
\newcommand{\bset}[1]{\left\{#1\right\}}
\newcommand{\cd}{\cdot}
\newcommand{\ld}{\ldots}

\newcommand{\g}{\tr{g}}
\newcommand{\I}{\tr{I}}
\newcommand{\IR}{\tr{IR}}
\newcommand{\CH}{\tr{CH}}

\newtheorem{proposition}{Proposition}

\begin{document}

\maketitle

\begin{abstract}
The throughput achievable in truncated Hybrid ARQ protocol (HARQ) using incremental redundancy (IR) in analyzed when transmitting over a block-fading channel whose state is unknown at the transmitter. We allow the transmission lengths to vary, optimize them efficiently via dynamic programming, and show that such a variable-rate HARQ-IR provides gains with respect to a fixed-rate transmission in terms of increased throughput and decreased average number of transmissions, reducing at the same time the outage probability.
\end{abstract}


\section{Introduction}\label{Sec:Introduction}

Automatic repeat request (ARQ) uses retransmissions to recover data lost due to errors inevitable when transmitting over variable and unreliable channels. ARQ is based on the principle that the receiver can inform the transmitter  about the transmission failure, to which the transmitter responds  retransmitting the lost date; ARQ used together with channel coding is known as hybrid ARQ (HARQ) \cite{Brayer68}. HARQ where we limit the number of allowed transmission attempts is known as truncated HARQ. 

In this work, we evaluate the throughput achievable in wireless links when using a truncated HARQ that conveys incremental,  redundancy (IR) in subsequent transmission attempts. For such a HARQ-IR system, we use random coding and maximum likelihood decoding assumptions of \cite{Caire01}\cite{Wu10}\cite{Tuninetti11}. We adopt the same simple scenario where each transmission attempt is carried out over independently fading channel and we generalize the assumptions of \cite{Caire01} allowing the transmission lengths (or -- rates) to vary throughout the transmissions attempts. We show how to efficiently find the  throughput-maximizing rates and we show gains obtained for a finite number of transmissions (truncated HARQ).

The idea of using variable-rate transmissions was already proposed and/or discussed in the literature but was not analyzed in the information-theoretic framework of \cite{Caire01}, which sets the upper bounds on the performance of any practical scheme. For example, a general formulation of the problem was provided in \cite{Visotsky03} which analyzed the infinite number of transmission attempts in abstraction of the channel model. The gains of variable-rate transmission over its fixed-rate counterpart for the predefined families of code were shown in \cite{Uhlemann03}\cite{Cheng03}\cite{LiuR03}\cite{Visotsky05}.  In \cite{Gopalakrishnan08b}\cite{Kim11} the correlated fading was considered, while \cite{Shen09} assumed that the channel stays constant for all transmission attempts. The idea of varying the transmission parameters appeared also in \cite{Tuninetti07}\cite{Gamal06}\cite{Nguyen10}\cite{Tuninetti11}, where power was varied on a per transmission-attempt basis. 

We are interested here in the practical case of truncated HARQ when the packet loss (outage) cannot be avoided. In such a case the throughput of HARQ may be optimized under constraints imposed on the outage probability \cite{Gopalakrishnan08b}\cite{Nguyen10} or without such constraints \cite{Tuninetti07}\cite{Uhlemann03}; the latter approach is also adopted in this paper.

In this work we analyze the ``conventional'' HARQ, i.e.,  when the return channel can carry only one-bit ACK/NACK messages \cite{Uhlemann03}\cite{Cheng03}\cite{Gopalakrishnan08b}. If, on the other hand, we allow the return channel to carry more bits, then, the parameters (rate or power) can be \emph{adapted} using such a  ``rich'' or ``multi-level'' feedback, e.g., \cite{Visotsky05}\cite{Gopalakrishnan08}\cite{Tuninetti11}. In the conventional' case, the \emph{adaptation} is not possible but the transmission parameters (rate or power) can be \emph{allocated}, that is, defined a priori for given channel conditions (e.g., the average SNR); this is focus of this work.

While power \emph{adaptation} improves the throughput \cite{Tuninetti11}, the power \emph{allocation} improves the diversity (asymptotic value of the outage for high SNR) but yields significant gains in terms of throughput only in the low-SNR range \cite{Tuninetti07}. Such conclusions resemble  those drawn in the context of adaptive modulation and coding \cite{Kim07} or in information-theoretic analysis of water-filling \cite{Goldsmith97_b}. In this work, interesting in medium-high SNR region, we assume a constant-power transmission as the gains obtained when allocating the power are often small \cite{Tuninetti07}.

The objective of this work is thus to evaluate the benefits of constant-power, variable-rate transmission for truncated HARQ when compared to the fixed-rate case analyzed in \cite{Caire01} and the main contributions are the following: a)~we show how to efficiently optimize the rates allocation for truncated HARQ with incremental redundancy, and b)~we asses the gains of variable-rate HARQ over its fixed-rate counterpart, showing that larger throughput, lower outage, and smaller average number of transmissions are yield.

\section{System Model}\label{Sec:Model}

In the transmission system under study, information bits are separated into packets of equal length of $N_\tr{b}$ bits, which are then  encoded into codeword of $N_\tr{s}$ complex symbols $x_{1}, x_{2},\ld, x_{N_\tr{s}}$ that are drawn randomly from the zero-mean complex Gaussian distribution with unitary variance. The symbols and gathered into $K$ sub-codewords $\mb{x}_{1}, \mb{x}_{2}, \ld, \mb{x}_{K}$ whose respective lengths are $N_{\tr{s},1}, N_{\tr{s},2},\ld, N_{\tr{s},K}$. We consider two ways of obtaining the sub-codewords:
\begin{enumerate}

\item A repetition coding (RTC), where the symbols are picked consecutively starting always with $x_{1}$
\begin{align}
    \mb{x}_{k}=[x_{1},\ld,x_{N_{\tr{s},k}}], \quad N_{\tr{s},k}\leq N_\tr{s}.
\end{align}
In this way, $\min_{k}\set{N_{\tr{s},k}}$ symbols are the same in the transmission attempts $1,\ld, k$.

\item An incremental redundancy (IR) transmission, where each sub-codewords is composed of different symbols
\begin{align}
    \mb{x}_{k}=[x_{t'_{k}+1},\ld,x_{t'_{k}+N_{\tr{s},k}}]\quad \tr{with}\quad t'_{k}=\sum_{l=1}^{k-1}N_{\tr{s},l}
\end{align}
This corresponds to puncturing of the codewords $\mb{x}=[x_{1},\ld, x_{N_\tr{s}}]$ into $K$ distinct sub-codewords $\mb{x}_{k}$ each of length $N_{\tr{s},k}, k=1,\ld, K$, where $\sum_{k=1}^{K}N_{\tr{s},k}=N_\tr{s}$ and $\mb{x}=[\mb{x}_{1},\ld, \mb{x}_K]$.  
\end{enumerate}

For convenience, we normalize the values of $N_{\tr{s},k}$ using $\rho_{k}=N_{\tr{s},k}/N_\tr{b}$, which has the meaning of the redundancy (measured by the number of channel uses per transmitted bit) and satisfy the relationship $\rho=N_\tr{s}/N_\tr{b}=\sum_{k=1}^{K}\rho_{k}$. We define also the rate of each transmission attempt $R_{k}=1/\rho_{k}$ and since the rate of the transmission attempts are not the same, we talk about variable-rate (VR) transmission, while if $\rho_{k}\equiv \rho_{1}, \forall k$ (or $R_{k}\equiv R_{1}$) we obtain the fixed-rate (FR) transmission considered before in \cite{Caire01} or \cite{Wu10}.

The ARQ process for each packet starts sending the sub-codeword $\mb{x}_{1}$. We assume that the feedback (or, \emph{return})  error-free channel exists, which allows the receiver to send (to the transmitter) a one-bit message required by the ARQ process (ACK or NACK). If the packet is not decoded  correctly\footnote{The receiver can determine if the decoding error occurs using an outer error-check code which causes the transmission overhead which we neglect for simplicity of the analysis.}, the NACK message is communicated by the receiver to the transmitter. Upon reception of a NACK message, knowing that the first sub-codeword was not decoded correctly, the transmitter sends a sub-codeword $\mb{x}_{2}$ composed of $N_{\tr{s},2}$ symbols. After unsuccessful decoding, another NACK message is generated to which the transmitter responds sending the codeword $\mb{x}_{3}$. This continues till the maximum allowed number of transmission attempts $K$ is reached (truncated HARQ) or until an ACK message, denoting a successful decoding, is received.

In a particular case of $\rho_{k}\equiv \rho_{1}$, the sub-codewords have the same length/rate and we recover the retransmission schemes analyzed in \cite{Caire01}. 

The channel remains constant during transmission of the $k$th sub-codeword $k=1,\ld, K$ and the received signal is given by 
\begin{align}
   \mb{y}_{k}=\sqrt{\gamma_{k}}\mb{x}_{k}+\mb{z}_{k} 
\end{align}
where $\mb{z}_{k}$ is the vector of zero-mean complex, unitary-variance uncorrelated Gaussian variables (modelling noise). The signal-to-noise ratio (SNR) $\gamma_{k}$ defines the channel state information (CSI) which is perfectly known/estimated at the receiver, but unknown to the transmitter. SNR does not change during the transmission of the sub-codeword but varies independently from one sub-codeword to another. This corresponds to a practical scenario where subsequent sub-codewords are not sent in adjacent time instants and, being sufficiently well separated, the realizations of the SNR become---to all practical extent---independent.  

The channel gains $\sqrt{\gamma}$ are Nakagami-$m$ distributed, so the SNR is characterized by the gamma function (PDF)
\begin{align}\label{PDF.channel}
\tr{p}(\gamma; m)=\frac{\gamma^{m-1}}{\Gamma(m)}\Bigl(\frac{m}{\ov{\gamma}}\Bigr)^m\exp\left(-\frac{m\gamma}{\ov{\gamma}}\,\right).
\end{align}
where $\ov{\gamma}$ is the average SNR. The cumulative density function of SNR is thus given by
\begin{align}\label{CDF}
\tr{F}(x; m)=\int_{0}^{x}\tr{p}(\gamma; m)\tr{d}\gamma=\Gamma(m, mx/\ov{\gamma}) 
\end{align}
with $\Gamma(m,\gamma)=\frac{1}{\Gamma(m)}\int_{0}^{\gamma}x^{m-1}\tr{e}^{-x}\tr{d}x$ and $\Gamma(m)=\Gamma(m,\infty)$ are, respectively, the incomplete gamma function and the gamma function.

The coding scheme is revealed to the transmitter, which in the $k$th transmission implements a maximum likelihood decoding using the observations $\tilde{\mb{y}}_{k}=[\mb{y}_{1},\ld,\mb{y}_{k}]$. 

The system-level implementation of the variable-rate HARQ described above deserves some comments. Namely, we may assume that each transmission contains only one sub-codeword in which case the duration of transmission attempts must vary. This might be a valid approach for a single-user communication where the transmitter and the receiver can negotiate the transmission time for each sub-codeword. On the other hand, it may be a questionable strategy in multi-user communications, where sharing the requirement for a variable-rate transmission with all the users is not practical. It might be possible to assign the resources (time) independently of the varying transmission length but it would lead to the bandwidth loss (sub-codewords shorter than the assigned transmission time slot) or to collisions (sub-codewords longer than the available time). 

\begin{figure}[tb]
\psfrag{xlabel}{$\ov{\gamma}$}
\psfrag{ylabel}[c]{Achievable rate}
\psfrag{P1}{$P_{1}$}\psfrag{P2}{$P_{2}$}\psfrag{P3}{$P_{3}$}\psfrag{P4}{$P_{4}$}
\psfrag{P5}{$P_{5}$}\psfrag{P6}{$P_{6}$}\psfrag{P7}{$P_{7}$}\psfrag{P8}{$P_{8}$}
\psfrag{P9}{$P_{9}$}\psfrag{P10}{$P_{10}$}\psfrag{P11}{$P_{11}$}\psfrag{P12}{$P_{12}$}
\psfrag{P13}{$P_{13}$}
\psfrag{G1}{$\gamma_{1}$}\psfrag{G2}{$\gamma_{2}$}\psfrag{G3}{$\gamma_{3}$}\psfrag{G4}{$\gamma_{4}$}
\psfrag{NF}{$N_\tr{F}$}
\begin{center}
\scalebox{1}{\includegraphics[width=0.8\linewidth]{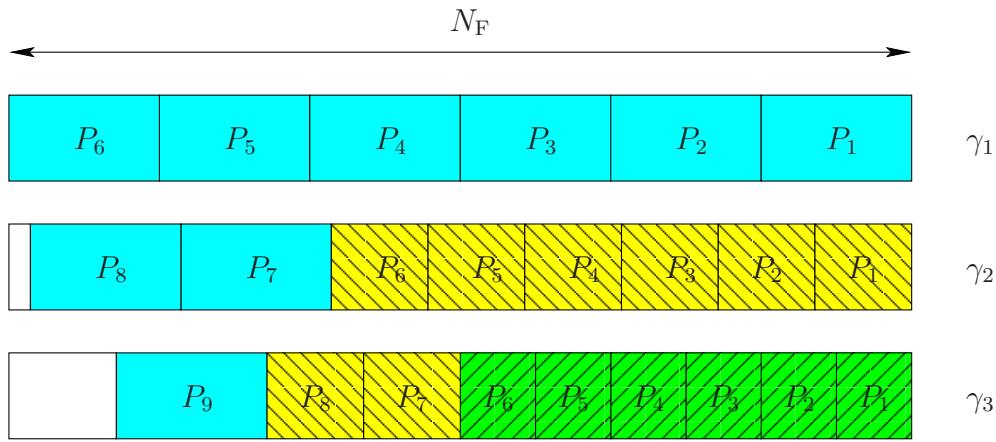}}
\caption{Example of the structure of three frames sent over channels with corresponding SNRs $\gamma_{1}$, $\gamma_{2}$, and $\gamma_{3}$ when delivering data packets  denoted by $P_{l}, l=1,\ld, 9$. The subcodewords having different lengths are identified with different colors and patterns. The first frame is filled up with subcodewords of length $N_{\tr{s},1}$ (thus, in our example, $N_\tr{F}=6 N_{\tr{s},1}$) corresponding to the packets $P_{1}-P_{6}$. When transmitting this frame with SNR $\gamma_{1}$, we assume $C(\gamma_{1})\rho_{1}<1$, consequently,  the decoder fails to decode the message in the packets $P_{1}-P_{6}$ and a NACK messages are sent to the transmitter. The next frame contains thus six subcodewords of length $N_{\tr{s},2}$ each carrying the redundancy for the undelivered packets and since, here, $N_{\tr{s},1}>N_{\tr{s},2}$, the ``empty'' space is filled with two subcodewords of the length $N_{\tr{s},1}$ corresponding of the packets  $P_{7}$ and $P_{8}$ that are ready for transmission. None of the packets is decoded after the transmission of the second frame so, again, six sub-codewords of length $N_{\tr{s},3}$, corresponding to the packets $P_{1}-P_{6}$ are sent as well as the sub-codewords of length $N_{\tr{s},2}$ corresponding to the packets $P_{7}$ and $P_{8}$. The residual time is filled with the sub-codeword corresponding to the packet $P_{9}$. Note, that the relative loss due to  unshaded/unfilled space can be made arbitrarily small loading the frame with many sub-codewords.}\label{Fig:frame}
\end{center}
\end{figure}

To avoid such a conceptual difficulty, we assume that the sub-codewords corresponding to different packets are gathered in frames that have the duration of $N_\tr{F}$ symbols. Such an assumption, also used in \cite{Liu04, Wang07} allows us to deal with variable-rate codewords to fill up the frame and  corresponds to TDMA-type communication, where users are provided with a fixed transmission time (frame). This is shown schematically in Fig.~\ref{Fig:frame}. We can easily see that the relative loss due to variable length of the sub-codewords can be made arbitrarily small, provided the number of packets in each frame is sufficiently large.

\section{Achievable throughput}

The definition of the throughput we use here follows \cite{Caire01}; according to  the \emph{reward-renewal} theorem \cite{Zorzi96} it is the ratio between the expected number of correctly received bits  (after up to $K$ transmissions) and the expected number of channel uses  $\ov{N}_{\tr{s}}$ required by the HARQ protocol to deliver the packet (in up to $K$ transmission attempts).

We denote by $\tr{NACK}_{k}$, the event of decoding failure in the $k$-th transmission and by  $f_{k}=\tr{Pr}\set{\tr{NACK}_{1}, \ld ,\tr{NACK}_{k-1}, \tr{NACK}_{k}}$ -- the probability of decoding failure after $k$ transmission attempts. The throughput can be then expressed as \cite{Visotsky03}\cite{Visotsky05}
\begin{align}
  \eta_{K}(\rho_{1},\ld, \rho_{K})&=\frac{1-f_{K}}{\rho_{1}+\sum_{k=2}^{K}f_{k-1}\rho_{k}}\label{TH.final}.
 \end{align}
which generalizes the results of \cite{Caire01} to the case of transmission with variable sub-codewords' lengths. Note that $f_{K}$ has the meaning of ``HARQ outage'', that is, the probability of loosing the data packet after the HARQ process is terminated.

The formulation \eqref{TH.final} is entirely general and depends only on the model of the channel and on the coding/decoding scheme. For example, it was used in \cite{Visotsky05} for convolutionally coded transmission while \cite{Caire01} used it in independently block-fading channel assuming that capacity-achieving codes are available  but under constraint $\rho_{k}\equiv\rho_{1}$. Here, we remove this constraint but still follow the approach of \cite{Caire01} that has the virtue of providing limits to any practical coding/decoding scheme. We thus assume that the coding/decoding scheme is ``capacity-achieving'' in the sense that the transmission is successful if the effective transmission rate is not greater than the accumulated mutual information between the sent and the received signals.\footnote{The existence of the codes satisfying this criterion when $N_\tr{b} \rightarrow \infty$ is discussed, e.g., in \cite{Malkamaki99}\cite{Gopalakrishnan08}.} This assumption as well as the way the transmitter/receiver deal with the retransmissions will affect the  variables $f_{k}$ used in \eqref{TH.final}. Namely, three HARQ schemes are considered:

\subsection{HARQ-I}\label{Sec:HARQ-I}

In HARQ type-I (HARQ-I), after $k$ transmissions, only the most recent received sub-codeword is used for decoding and others are discarded (in \cite{Caire01} this scheme was denoted as ALO). In such a case, the decoding failures are independent of each others and the probability of losing a packet after $ k$ transmissions is calculated as \cite{Caire01}
\begin{align}\label{PLR}
  f_{\tr{I},k}=\prod_{l=1}^{k}\tr{Pr}\bset{ C(\gamma_{l})\rho_{l}< 1  } = \prod_{l=1}^{k} \nu( \rho_{l} )
\end{align}
where where 
$  C(\gamma)=\log_{2}(1+\gamma)$
is the average mutual information (per channel use) when transmitting with SNR $\gamma$ and $\nu(\rho)=\tr{F}(2^{1/\rho}-1; m)$ is the probability of outage (after a single transmission) when transmitting with redundancy $\rho$. 

The throughput of HARQ-I is then given by 
\begin{align}\label{TH.HARQ-I}
     \eta_{\I,K}\bigl(\rho_{1},\ld, \rho_{K}\bigr)= \frac{1- \prod_{k=1}^{K}\nu(\rho_{k})} { \rho_{1}+\sum_{k=2}^{K}\rho_{k}\prod_{l=1}^{k-1}\nu(\rho_{l}) }
\end{align}
and the optimal throughput is denoted as $\hat{\eta}_{\I,K}=\max_{\rho_{1},\ld, \rho_{K}}\eta_{\I,K}\bigl(\rho_{1},\ld, \rho_{K}\bigr)$.

\begin{proposition}\label{Prop:HARQ-I}
  The maximal throughput of HARQ-I $\hat{\eta}_{\I,K}$ is independent of $K$, i.e.,  $\hat{\eta}_{\tr{I},K}\equiv \hat{\eta}_{\tr{I}}=\eta_{\I,1}(\hat{\rho}_{\I})$ where $\hat{\rho}_{\I}=\arg_{\rho}\max \frac{1-\nu(\rho)}{\rho}$, and is yield with fixed-rate HARQ (FR-HARQ-I) $\hat{\rho}_{\tr{I},l}=\hat{\rho}_{\I}, l=1,\ld,K$. 
\IEEEproof 
Since $\hat{\eta}_{\I,1}\geq\frac{1-\nu(\rho)}{\rho}$, where the equality  hold only for $\rho=\hat{\rho}_{\I}$,  we can use 
$\rho_{k}\geq\frac{1-\nu(\rho_{k})}{\hat{\eta}_{\I,1}}$ in \eqref{TH.HARQ-I}, which yields the following inequality
\begin{align}
\eta_{\I,K}(\rho_{1},\ld,\rho_{K})&\leq\hat{\eta}_{\I,1} \frac{1- \prod_{k=1}^{K}\nu(\rho_{k})} {1-\nu( \rho_{1})+\sum_{k=2}^{K}(1-\nu(\rho_{k}))\prod_{l=1}^{k-1}\nu(\rho_{l}) }\nonumber\\
&=\hat{\eta}_{\I,1}\frac{1- \prod_{k=1}^{K}\nu(\rho_{k})} {1-\nu( \rho_{1})+\sum_{k=1}^{K-1}\prod_{l=1}^{k}\nu(\rho_{l}) -\sum_{k=2}^{K}\prod_{l=1}^{k}\nu(\rho_{l}) }\nonumber\\
\eta_{\I,K}(\rho_{1},\ld,\rho_{K}) &\leq \hat{\eta}_{\I,1}=\hat{\eta}_{\I}
\end{align}
thus $\hat{\eta}_{\I}$ is the maximum throughout of VR-HARQ-I, achievable only if $\rho_{k}=\hat{\rho}_{\I}, k=1,\ld,K$.
\end{proposition}

According to Proposition~\ref{Prop:HARQ-I}, the fixed-rate HARQ-I is optimal so the same sub-codeword may be used for each transmission and the transmitter can apply the RTC transmission scheme defined in Sec.~\ref{Sec:Model}.

Proposition 1 that is valid for any $K$ may be seen as a generalization of Corrolary~1 in \cite{Visotsky03} valid for  $K\rightarrow\infty$.


\subsection{HARQ-IR}\label{Sec:IR}

In incremental redundancy HARQ (HARQ-IR) the transmitted sub-codewords are obtained according to IR principle described in Sec.~\ref{Sec:Model} and the decoding fails in the $k$-th transmission attempt if the accumulated mutual information is lower than the transmission rate, which yields the following condition \cite{Gopalakrishnan08}
\begin{align}\label{PLR.IR}
  f_{\tr{IR},k}=  \tr{Pr}\Bigl\{ \sum_{l=1}^{k} {C(\gamma_{l})}{ \rho_{l}}  < 1 \Bigr\}.
\end{align}
where $\gamma_{l}$ is the SNR during $l$th transmission attempt. 


To calculate $f_{\tr{IR},k}$ we may proceed as suggested in \cite{Caire01} introducing random variable $v_{l}=C(\gamma_{l})\cd\rho_{l}$, $l=1,\ld, k$ whose PDF can be obtained by definition as $\g_{l}(x)=\ln(2)\cd\tr{p}( 2^{x/\rho_{l}}-1; m)2^{x /\rho_{l}}/\rho_{l}$. This is what will be called the ``exact'' calculation. 

Alternatively, we may approximate $v_{l}$ by a Gaussian variable,\cite{Wu10}, i.e.,
\begin{align}\label{Gauss.approx.pdf}
 \g_{l}(x) \approx \tilde{\g}_{k}(x)=\frac{1}{\sqrt{2\pi}\rho_{l}\sigma_{m}}\exp\Bigl(-\frac{(x-\ov{C}_{m}\rho_{l})^{2}}{2\rho^{2}_{l}\sigma_{m}^{2}}\Bigr)
\end{align} 
where
\begin{align}
\ov{C}_{m}&=\int_{0}^{\infty}C(\gamma)\tr{p}(\gamma; m)\tr{d}\gamma\label{C.ergodic}\\
\sigma^{2}_{m}&=\int_{0}^{\infty}C^{2}(\gamma)\tr{p}(\gamma; m)\tr{d}\gamma-\ov{C}_{m}^{2}\label{var.ergodic}
\end{align}
are, respectively  the mean of $C(\gamma)$ (i.e., the ergodic capacity), and the variance of $C(\gamma)$.

Since $v_{l}, l=1,\ld, k$ are independent, $f_{\tr{IR},k}=\tr{Pr}\left\{ \sum_{l=1}^{k} v_{l}<1   \right\}=  \int_{0}^{1} \ov{\g}_{k} (x)\tr{d} x$,  where $\ov{\g}_{k}(x)$ is a convolution of $\g_{l}(x)$. The latter must be calculated numerically, e.g., via direct/inverse Fourier transform if the exact form of $\g_{l}(x)$ is used, while, applying \eqref{Gauss.approx.pdf} we obtain a closed-form approximation of \eqref{PLR.IR}
\begin{align}\label{Gauss.approx.f}
  f_{\tr{IR},k}\approx \tilde{f}_{\tr{IR},k}= Q\left( \xi \frac{{X_{k}-1}}{Y_{k}} \right),
\end{align}
where $Q(x)=\frac{1}{\sqrt{2\pi}}\int_{x}^{\infty}\exp(-t^{2}/2)\tr{d}t$, $\xi=\frac{\ov{C}}{\sigma_{C}}$, $X_{k}=\sum_{l=1}^{k}\rho'_{l}$, $Y_{k}=\sqrt{\sum_{l=1}^{k}{\rho'}^{2}_{l}}$, and $\rho'=\rho\cd\ov{C}$. 

\begin{proposition}\label{Prop:HARQ-IR.bounds}
Denoting by $\tilde{\eta}_{\tr{IR},K}(\rho_{1},\ld, \rho_{K})$ the approximation of the throughput obtained using \eqref{Gauss.approx.f} in \eqref{TH.final}, the following inequality holds
\begin{align}
\tilde{\eta}_{\IR,K}(\rho_{1},\ld, \rho_{K})&\geq\check{\eta}_{\IR,K}(\rho_{1},\ld, \rho_{K}) = \ov{C}\frac{1-\check{f}_{\tr{IR},K}}{\rho'_{1}+\sum_{k=2}^{K}\rho'_{k}\cd \check{f}_{\tr{IR},k-1}}\label{TH.IR.Lbound}
\end{align}
where $\check{f}_{\tr{IR},k}=Q\bigl(\xi(1-1/X_{k})\bigr)$.
\IEEEproof The obvious relationship $Y^{2}_{k} \leq X^{2}_{k}$ used in \eqref{Gauss.approx.f} yields $\tilde{f}_{\IR,k}\leq\check{f}_{\tr{IR},k}$. From this inequality and knowing that the throughput decreases monothonicaly with $f_{k}$ (if $\rho_{k}$ are kept constant), we immediately obtain the lower bound  \eqref{TH.IR.Lbound}.
\end{proposition}

The bound \eqref{TH.IR.Lbound} will be useful to optimize the throughput in Sec.~\ref{Sec:Optimization}.

\subsection{HARQ-CHASE}\label{Sec:CH}

Instead of discarding the packets that were not decoded correctly (as done in HARQ-I), the receiver should take advantage of all received packets and if  RTC is employed, the received signals should be weighted by the corresponding SNR and added up. This is known as maximum ratio combining (MRC) or Chase combining \cite{Cheng06}. Then, the decoder used the following signal 
\begin{align}\label{combine.Chase}
  \tilde{\mb{y}}_{k}=\sum_{l=1}^{k}\sqrt{\gamma_{l}}\cd\mb{y}'_{l}
\end{align}
where $\mb{y}'_{k}=[\mb{y}_{k},0, 0, \ld, 0]$ are zero-padded version of the received signal $\mb{y}_{k}$. The padding is used to make the notation compact and may be seen as an operation carried out at the receiver thus it does not affects the throughput. 

Although, in the case of a fixed-rate HARQ, it was shown to introduce little gain over HARQ-I \cite{Caire01, Tuninetti07}, Chase-combining is the most the receiver can do when RTC is implemented at the transmitter so we deal with this case for completeness of our analysis.



Calculation of the decoding failure probability is slightly more involved in this case.

First, for convenience, we reorder the variables $\gamma_{1}, \ld, \gamma_{k}$ so that the corresponding sub-codewords lengths' after reordering are non-decreasing  $\rho_{\kappa_{ 1}} \leq \rho_{\kappa_{2}} \leq \ld  \leq \rho_{\kappa_{k} }$, where $\kappa_{1},\ld, \kappa_{k}$ is a permutation of $1,\ld, k$.  We emphasize that the reordering is merely a concept simplifying the analysis and not an a priori constraints on the values $\rho_{k}$. 

\begin{figure}[tb]
\psfrag{e1}{$\bs{\epsilon}_{k,1}$}\psfrag{e2}{$\bs{\epsilon}_{k,2}$}\psfrag{ek}{$\bs{\epsilon}_{k,k}$}
\psfrag{LD}{$\ld$}
\psfrag{X1}{$\bs{x}_{\kappa_1}$}\psfrag{X2}{$\bs{x}_{\kappa_2}$}\psfrag{Xk-1}{$\bs{x}_{\kappa_{k-1}}$}\psfrag{Xk}{$\bs{x}_{\kappa_k}$}
\psfrag{G1}{$\gamma_{\kappa_1}$}\psfrag{G2}{$\gamma_{\kappa_2}$}\psfrag{Gk-1}{$\gamma_{\kappa_{k-1}}$}\psfrag{Gk}{$\gamma_{\kappa_k}$}
\psfrag{Ns}{$N_{\tr{s},k}$}
\begin{center}
\scalebox{1}{\includegraphics[width=0.8\linewidth]{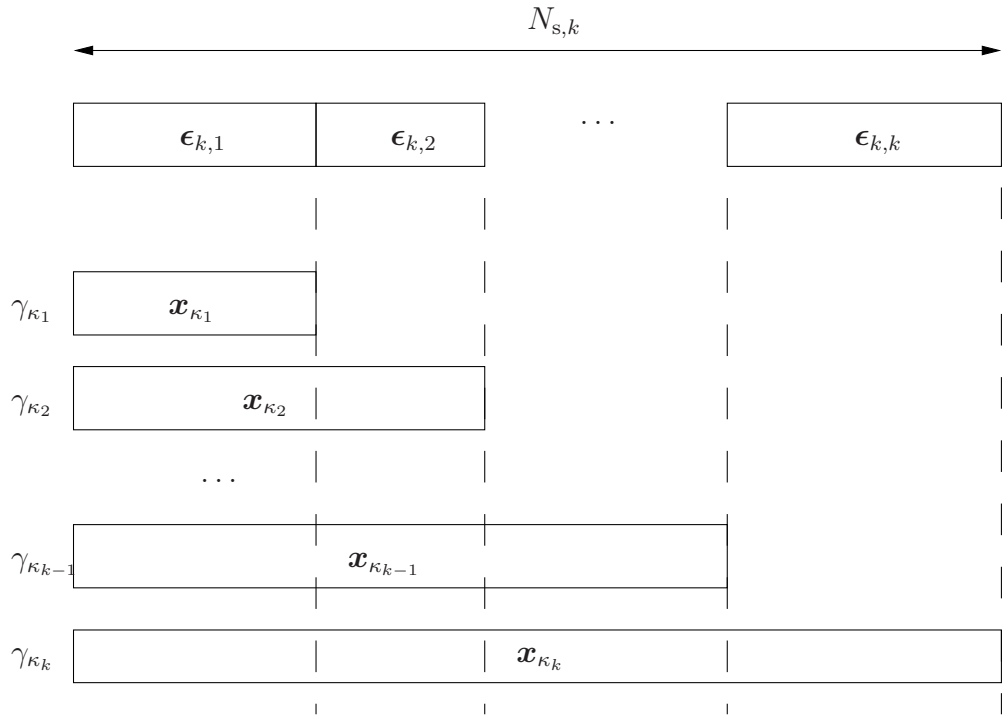}}
\caption{In RTC, the parts of the transmitted sub-codewords that contain the same symbols are identified as ``chunks'' $\bs{\epsilon}_{k,l}$; symbols in each chunk experience the same equivalent SNR $\tilde{\gamma}_{k,l}=\sum_{f=l}^{k}\gamma_{\kappa_{f}}$.}\label{Fig:chunks}
\end{center}
\end{figure}

As illustrated in Fig.~\ref{Fig:chunks}, thanks to the reordering we are able to identify $k$ ``chunks'' of the sub-codewords
\begin{align}\label{eps.x}
\bs{\epsilon}_{k,1}&=  [x_{1}, \ld, x_{N_{\tr{s},\kappa_{1}}}],\nonumber\\
\bs{\epsilon}_{k,l}&=  [x_{N_{\tr{s},\kappa_{l-1}}+1}, \ld, x_{N_{\tr{s},\kappa_{l}}}], \quad l=2,\ld, k,\nonumber
\end{align}
each with normalized redundancy $\tilde{\rho}_{l} =\rho_{\kappa_{l}}-\rho_{\kappa_{l-1}}$  (we set $\rho_{\kappa_{0}}\equiv 0)$, such that all  symbols in the chunk $\bs{\epsilon}_{k,l}$ were transmitted in the transmissions attempts indexed with $\kappa_{l},\kappa_{l+1},\ld, \kappa_{k}$. 

After simple algebra, the combining of the received signals \eqref{combine.Chase} yields 
\begin{align}
  \tilde{\mb{y}}_{k}   &=[\tilde{\mb{y}}_{k,1}, \tilde{\mb{y}}_{k,2}, \ld,\tilde{\mb{y}}_{k,k}]\\
  \tilde{\mb{y}}_{k,l} &=\sqrt{\tilde{\gamma}_{k,l}}\left[\sqrt{\tilde{\gamma}_{k,l}}\cd\bs{\epsilon}_{k,l} + \bs{\xi}_{k,l}\right]
\end{align}
where 
\begin{align}
\tilde{\gamma}_{k,l} =\sum_{f=l}^{k}\gamma_{\kappa_{f}}.
\end{align}
is the equivalent SNR for the chunk $\bs{\epsilon}_{k,l}$ and $\bs{\xi}_{k,l}$ is a zero-mean,  unitary-variance Gaussian vector modelling ``equivalent noise'' affecting the chunk $\bs{\epsilon}_{k,l}$.

Since the symbols in the chunks $\bs{\epsilon}_{k,l}$ are mutually independent, the parts $\tilde{\mb{y}}_{k,l}$ of the received signal $\tilde{\mb{y}}_{k}$  may be seen as the result of transmission of the chunks $\bs{\epsilon}_{k,l}$ over the channel with SNR $\tilde{\gamma}_{k,l}$. Consequently, Chase combining may be seen as a form of IR transmission with redundancy  $\tilde{\rho}_{l} $ and  the probability of the decoding failure is given by  
\begin{align}
  f_{\CH,k}&=  \tr{Pr}\Bigl\{ \sum_{l=1}^{k} {C(\tilde{\gamma}_{k,l})}{ \tilde{\rho}_{l}}  < 1 \Bigr\}
  \label{PLR.CH}
\end{align}
that,  in  the case of a fixed-rate transmission, boils down to the formula shown in \cite{Caire01}. Namely, since in fixed-rate HARQ-CHASE $\tilde{\rho}_{1}=\rho_{1}$, $\tilde{\rho}_{l}=0, l=2,\ld,k$, and $\tilde{\gamma}_{k,1}=\sum_{l=1}^{k}\gamma_{k}$, then the decoding failure is calculated in a closed form 
\begin{align}
f_{\CH,k}=\tr{Pr}\bset{C\bigl(\sum_{l=1}^{k}\gamma_{k}\bigr)\cd \rho_{1}<1}
=\tr{F}(2^{1/\rho_{1}}-1; m\cd k).
\end{align}

While \eqref{PLR.CH} resembles \eqref{PLR.IR}, the equivalent SNRs $\tilde{\gamma}_{k,l}$ appearing in \eqref{PLR.CH} are not independent (unlike in the case of HARQ-IR), so the approach of Sec.~\ref{Sec:IR}, based on the convolution of the individual PDFs cannot be applied and a multidimensional integration over $\gamma_{1},\ld, \gamma_{k}$ is required  
\begin{align}
  f_{\CH,k}&=\int_{\mc{D}_{k}}\prod_{l=1}^{k} \tr{p}(\gamma_{l}; m)\tr{d}\gamma_{1}\ld\tr{d}\gamma_{k}\label{PLR.CH.Num}
\end{align}
where $\mc{D}_{k}=\set{ \gamma_{1},\ld,\gamma_{k} : \sum_{l=1}^{k} {C(\tilde{\gamma}_{k,l})}{ \tilde{\rho}_{l}}<1 }$, so
\begin{align}
f_{\CH,k}&=\int_{0}^{z_{1} }\tr{p}(\gamma_{1}; m)\tr{d}\gamma_{1}\ld\int_{0}^{z_{k-1}}\tr{p}(\gamma_{k-1}; m)\tr{d}\gamma_{k-1} \int_{0}^{z_{k} }\tr{p}(\gamma_{k}; m)\tr{d}\gamma_{k},\label{PLR.CH.Num2}
\end{align}
where the integration limit for the SNR $\gamma_{l}$ depends on the values taken by the SNRs $\gamma_{l+1}, \ld, \gamma_{k}$
\begin{align}
  z_{l}\equiv z_{l}(\gamma_{l+1}, \ld, \gamma_{k})=\left[2^{R_{l} (1-\sum_{f=l+1}^{k}\frac{1}{R_{l}}[C(\tilde{\gamma}_{f})-C(\tilde{\gamma}_{f+1})]  )  }-1\right](1+\tilde{\gamma}_{l+1}).
\end{align}

To implement \eqref{PLR.CH.Num2} we used the Gauss-Laguerre formulae with 10 (for $m=1,2$) or 40 (for $m=\frac{1}{2}$) points in each of $k$ dimensions of $\mc{D}_{k}$.

The multidimensional calculation was particularly computationally-intensive  for $K>4$ and the results do not seem very relevant beyond this point as virtually all improvement is due to the second transmission attempt.


\subsection{Limiting cases}\label{Sec:Bounds}

We know from \cite{Caire01} that for a fixed-rate HARQ-IR
\begin{align}\label{C.ergodic}
   \hat{\eta}_{\IR,K} \xrightarrow[K \rightarrow \infty]{}  \ov{C}_{m}
\end{align}
where $\ov{C}_{m}$ is the ergodic capacity, defined in \eqref{C.ergodic}.

We also known that, for a given set of $\rho_{1}, \ld, \rho_{K}$, the relationship $f_{\I,k}> f_{\CH,k} > f_{\IR,k}$  holds for all $k$ \cite{Caire01}, and since, for the given $\rho_{1},\ld, \rho_{K}$, the throughput $\eta_{K}$ \eqref{TH.final} monotonically decreases when $f_{k}$ increases, we  conclude that for $K<\infty$
\begin{align}
\hat{\eta}_{\I,1}=\hat{\eta}_{\I,K} <\hat{\eta}_{\CH,K} < \hat{\eta}_{\IR, K}<\ov{C}.\label{TH.relationship}
\end{align}
Thus,  the throughput of HARQ schemes with fixed-power transmission, operating without knowledge of instantaneous SNR, is lower-bounded by $\hat{\eta}_{\I}$ defined in Sec.~\ref{Sec:HARQ-I} and upper-bounded by the ergodic capacity $\ov{C}$.

The limiting case $K\rightarrow\infty$ is also interesting since, as stated in \cite[Lemma~1]{Visotsky03}, when the receiver does not discard packets (as it is the case for HARQ-IR and HARQ-CHASE), the optimal redundancy sequence must be non-increasing, i.e.,  $\rho_{\IR,k}\geq\rho_{\IR,k+1}$ and $\rho_{\CH,k}\geq\rho_{\CH,k+1}$.

\section{Optimization}\label{Sec:Optimization}

The ``design'' of the HARQ scheme consists in the maximization of the throughput over the redundancy values $\rho_{1},\ld,\rho_{K}$. In the case of FR-HARQ the exhaustive search over one-dimensional space is relatively simple. On the other hand, the solutions for VR-HARQ-IR and VR-HARQ-CHASE are more difficult to find as their require a multidimensional optimization. 

To maximize \eqref{TH.final} we might use a gradient-based method but the initialization of the variables is critical to ensure rapid convergence and to avoid getting trapped far from the global optimum (both - not guaranteed in non-concave functions we deal with, cf.~\cite[Fig.~1]{Szczecinski10}) , so we used this approach only in VR-HARQ-CHASE where various initializations were tested and the solutions were compared to the random initializations. This was tedious but feasible as it was done only for $K\leq 4$.

In case of VR-HARQ-IR, different approach was adopted: instead of maximizing the throughput $\eta_{\IR,K}$ we maximizing the lower bound \eqref{TH.IR.Lbound}. The problem is greatly simplified since each term $\check{f}_{\IR,k}$ depends uniquely on $X_{k}=\sum_{l=1}^{k}\rho'_{l}$ and the optimization may be written as 
\begin{align}\label{max.bound}
   \max_{\rho_{1},\ld,\rho_{K}} \check{\eta}_\tr{IR}(\rho_{1},\ld,\rho_{K}) = \max_{X} \frac{1-f_{K}(X)}{V_{K}(X)} 
\end{align}
where $f_{k}(X)= \check{f}_{\IR,k}=Q\bigl(\xi  \sqrt{k}(1-1/X)\bigr)$ and
\begin{align}
  V_{k}(X)&=\min_{\substack{\rho'_{1},\ld,\rho'_{k}:\\\sum_{l=1}^{k}\rho'_{l}=X}}  \rho'_{1}+\sum_{l=2}^{k} \rho'_{l}f_{l-1}(X_{l-1})\\
                &=\min_{0\leq\rho'_{k}\leq X}  \min_{\substack{\rho'_{1},\ld,\rho'_{k-1}:\\\sum_{l=1}^{k-1}\rho'_{l}=X-\rho'_{k}}}  \rho'_{1} + \sum_{l=2}^{k-1} \rho'_{l}f_{l-1}(X_{l-1})     +    \rho'_{k} f_{k-1} (X-\rho_{k})\\
                &=\min_{0\leq \rho \leq X}   V_{k-1}(X-\rho) + \rho f_{k-1}(X-\rho)\label{min.Vk}.
\end{align}

For a given $X$, the minimization in \eqref{min.Vk} is done over one variable ($\rho=\rho_{k}$) provided the results of the minimization $V_{k-1}(X)$ are known for all arguments $X$. That is, first we solve $V_{2}(X)=\min_{\rho} \set{X-\rho + \rho f_{1}(X-\rho)} $, next $V_{3}(X)=\min_{\rho} \set{V_{2}(X-\rho) + \rho f_{1}(X-\rho)} $, etc. This recursive formulation is characteristic of the so-called dynamic programming (DP) \cite{Bertsekas05_book} whose application for throughput optimization  was already suggested in \cite{Visotsky03}. The direct benefit is that the optimization \eqref{max.bound} over $K$-dimensions is reduced to $K$, one-dimensional functional optimizations, which greatly simplifies the implementation. 

The function $V_{k}(X)$ is not obtained in the closed-form, so we discretized $X$ using 50-100 points over the domain $X \in(0, k)$, where the bounding of $X$ by $k$ is not restrictive and comes from the heuristic observation that $\rho'_{k}<1$, i.e., each rate $R_{k}=1/\rho_{k}$ is greater than the ergodic capacity $\ov{C}$.

The optimization results are stored as $\rho_{k}(X)=\arg\min_{\rho}  V_{k-1}(X-\rho) + \rho f_{k-1}(X-\rho)$, so once the functions $V_{k}(X)$ are obtained, we can recover the solution $\hat{\rho'}_{k}$ that maximizes the bound:
\begin{align}
   \hat{\rho'}_{k}=\rho_{k}(\hat{X}_{k})
\end{align}
where   $\hat{X}_{K}=\arg_{X}\max \frac{1-f_{K}(X)}{V_{K}(X)}$ and $\hat{X}_{k-1}=\hat{X}_{k}-\rho_{k}(\hat{X}_{k})$.

We note that while the approximate expressions for $f_{\IR,k}$ and $\check{\eta}_{\IR,K}$ are used in DP optimization, the throughput values we show in the following are based on the exact calculation of $f_{\IR,k}$. We also verified that using the DP-based results as the initialization to the gradient-based optimization  yields practically the same values of the throughput as those we show.

\section{Numerical Results}

\begin{figure}[tb]
\psfrag{xlabel}{$\ov{\gamma}$}
\psfrag{ylabel}[c]{$\eta$}
\psfrag{XXXXXXXXXXXXXXX1}{VR-HARQ, $K=2$}
\psfrag{XXXXXXXXXXXXXXX2}{VR-HARQ, $K=4$}
\psfrag{XXXXXXXXXXXXXXX3}{VR-HARQ, $K=8$}
\psfrag{XXXXXXXXXXXXXXX4}{FR-HARQ, $K=2$}
\psfrag{XXXXXXXXXXXXXXX5}{FR-HARQ, $K=4$}
\psfrag{XXXXXXXXXXXXXXX6}{FR-HARQ, $K=8$}
\psfrag{YYYYYYYYYYYYYYY1}{VR-HARQ-IR}
\psfrag{YYYYYYYYYYYYYYY2}{FR-HARQ-IR}
\psfrag{YYYYYYYYYYYYYYY3}{VR-HARQ-CHASE}
\psfrag{YYYYYYYYYYYYYYY4}{FR-HARQ-CHASE}
\psfrag{YYYYYYYYYYYYYYY5}{$K=1$}
\psfrag{YYYYYYYYYYYYYYY6}{$\ov{C}$}

\begin{center}
\scalebox{1}{\includegraphics[width=1\linewidth]{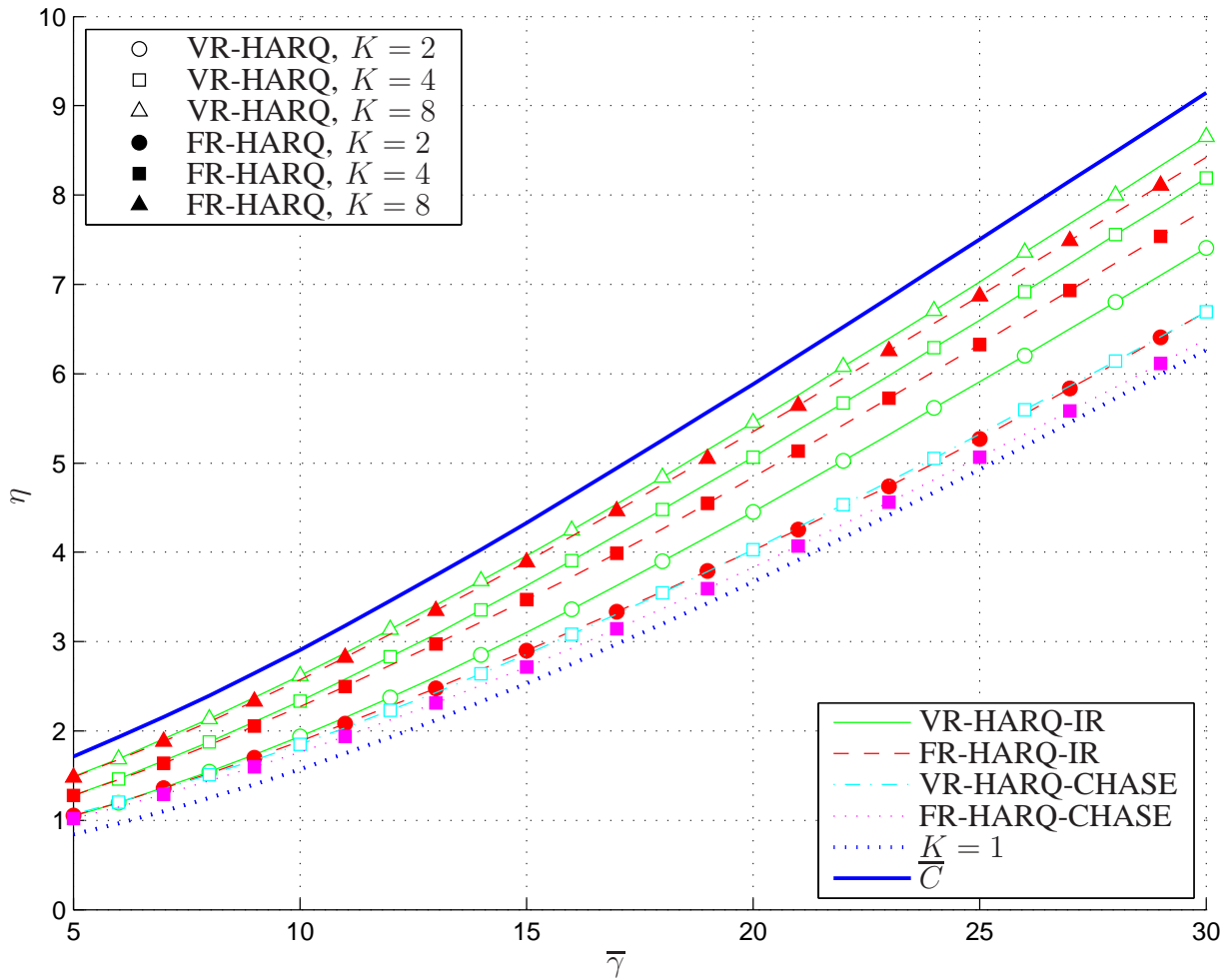}}
\caption{Throughput in a block-fading Rayleigh channel for VR-HARQ-IR (solid, green line), FR-HARQ-IR (dashed, red line) and $K=2, 4, 8$ as well as VR-HARQ-CHASE (dot-dashed, cyan line) and FR-HARQ-CHASE (dotted, magenta line) shown only for $K=4$. The upper bound (thick, solid, blue line) corresponds to the ergodic capacity $\ov{C}$ while the lower one (dotted, blue line) corresponds to transmission without HARQ ($K=1$). That results of VR-HARQ-CHASE ($K=4$) and VR-HARQ-IR ( $K=2$) are practically superimposed, as well as are those of FR-HARQ-CHASE ($K=4$) and of transmission without HARQ. }\label{Fig:TH}
\end{center}
\end{figure}

The optimized throughput of fixed- and variable-rate HARQ-IR is shown in Fig.~\ref{Fig:TH} for $K=2, 4, 8$ for Rayleigh fading channel (i.e. with $m=1$), where the gain due to variable-rate transmission is particularly notable for HARQ-IR while it is very slight when considering HARQ-CHASE, which at best (with $K=4$) equals the performance of FR-HARQ-IR with $K=2$.

\begin{figure}[tb]
\psfrag{xlabel}{$K$}
\psfrag{ylabel}{$\chi$}
\psfrag{XXXXX1}{$m=\frac{1}{2}$}
\psfrag{XXXXX2}{$m=1$}
\psfrag{XXXXX3}{$m=2$}
\psfrag{CHASE}{CHASE}\psfrag{IR}{IR}
\begin{center}
\scalebox{1}{\includegraphics[width=1\linewidth]{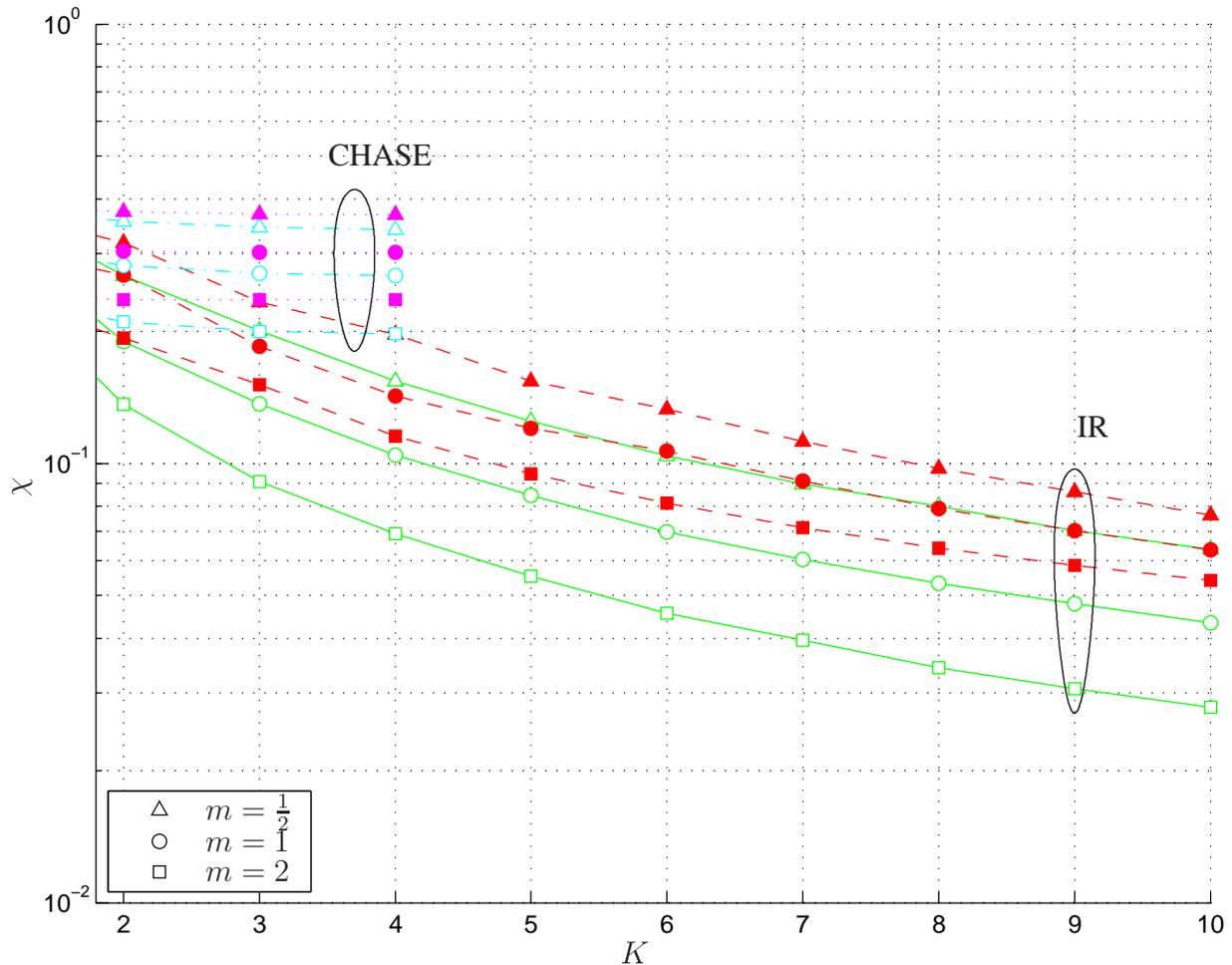}}
\caption{Residual throughput $\xi=1-\eta/\ov{C}$ for VR-HARQ-IR (solid, green line), FR-HARQ-IR (dashed, red line), FR-HARQ-CHASE (dotted, blue line), and VR-HARQ-CHASE (dashed-dotted, magenta line) is shown for varying $K$ and Nakagami-$m$ fading with $m=\frac{1}{2}, 1, 2$ and $\ov{\gamma}=30$dB. For FR/VR- HARQ-CHASE the results up to $K=4$ are shown due to high computation load of the throughput calculation.}\label{Fig:convergence}
\end{center}
\end{figure}

The gain in terms of throughput offered by VR-HARQ-IR is particularly clear for $K=2$ and to complement the results of Fig.~\ref{Fig:TH}, we evaluate and show in Fig.~\ref{Fig:convergence} the ``residual throughput'' 
\begin{align}
   \chi=1-\frac{\eta_\tr{IR}}{\ov{C}}
\end{align}
i.e., the relative gap between the throughput attained with up to $K$ transmissions and the maximum achievable throughput (ergodic capacity).  The relative gain of the VR-HARQ  with respect to FR-HARQ remain roughly constant for all $K$ but of course the absolute difference diminishes -- as expected -- with $K$ since, asymptotically both schemes are equivalent. These gains are also more notable when increasing $m$. The ``saturation'' of the throughput of HARQ-CHASE scheme is also clearly shown.

\begin{figure}[tb]
\psfrag{xlabel}{$k$}
\psfrag{ylabel}{$\rho_{k}\cd\ov{C}$}
\psfrag{XXXXXXXX1}{$\ov{\gamma}=$10dB}
\psfrag{XXXXXXXX2}{$\ov{\gamma}=$20dB}
\psfrag{VR-HARQ-IR}{VR-HARQ-IR}
\psfrag{FR-HARQ-IR}{FR-HARQ-IR}
\psfrag{VR-HARQ-CHASE}{VR-HARQ-CHASE}
\psfrag{FR-HARQ-CHASE}{FR-HARQ-CHASE}
\scalebox{1}{\includegraphics[width=1\linewidth]{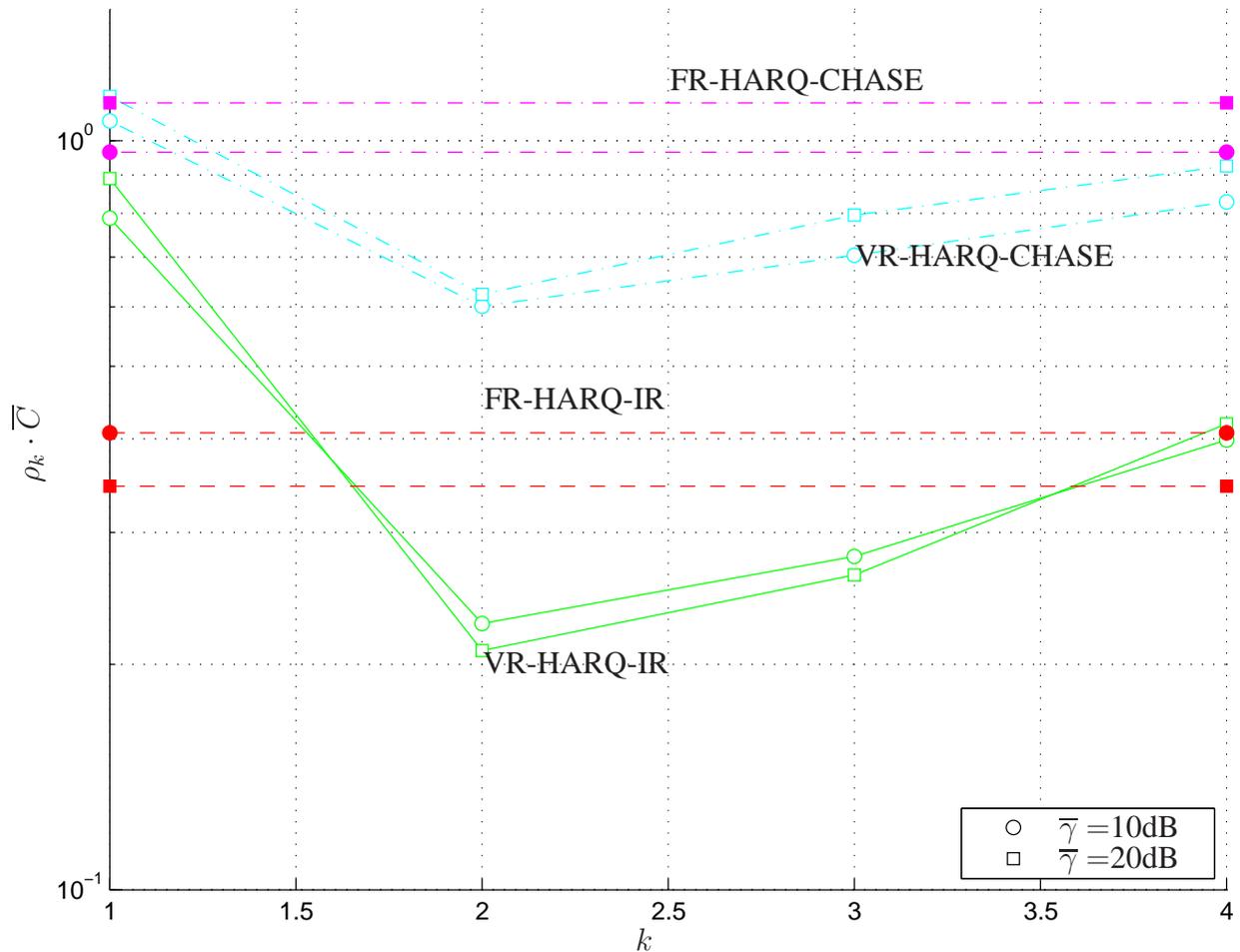}}
\caption{Throughput-maximizing normalized redundancy $\rho'_{k}=\rho\cd\ov{C}, k=1,\ld,K$ ($K=4$) for VR-HARQ-IR (solid, green line), FR-HARQ-IR (dashed, red line), VR-HARQ-CHASE (dashed-dotted, cyan line), and FR-HARQ-CHASE (dashed-dotted, magenta line); for $\ov{\gamma}=$10dB and $\ov{\gamma}=$20dB.}\label{Fig:rates}
\end{figure}

\begin{figure}[tb]
\psfrag{xlabel}{$k$}
\psfrag{ylabel}{$\rho_{k}\cd\ov{C}$}
\psfrag{XXXXX1}{$K=2$}
\psfrag{XXXXX2}{$K=4$}
\psfrag{XXXXX3}{$K=6$}
\psfrag{XXXXX4}{$K=8$}
\psfrag{XXXXX5}{$K=10$}
\psfrag{YYYYYYYYYY1}{VR-HARQ-IR}
\psfrag{YYYYYYYYYY2}{FR-HARQ-IR}
\begin{center}
\scalebox{1}{\includegraphics[width=1\linewidth]{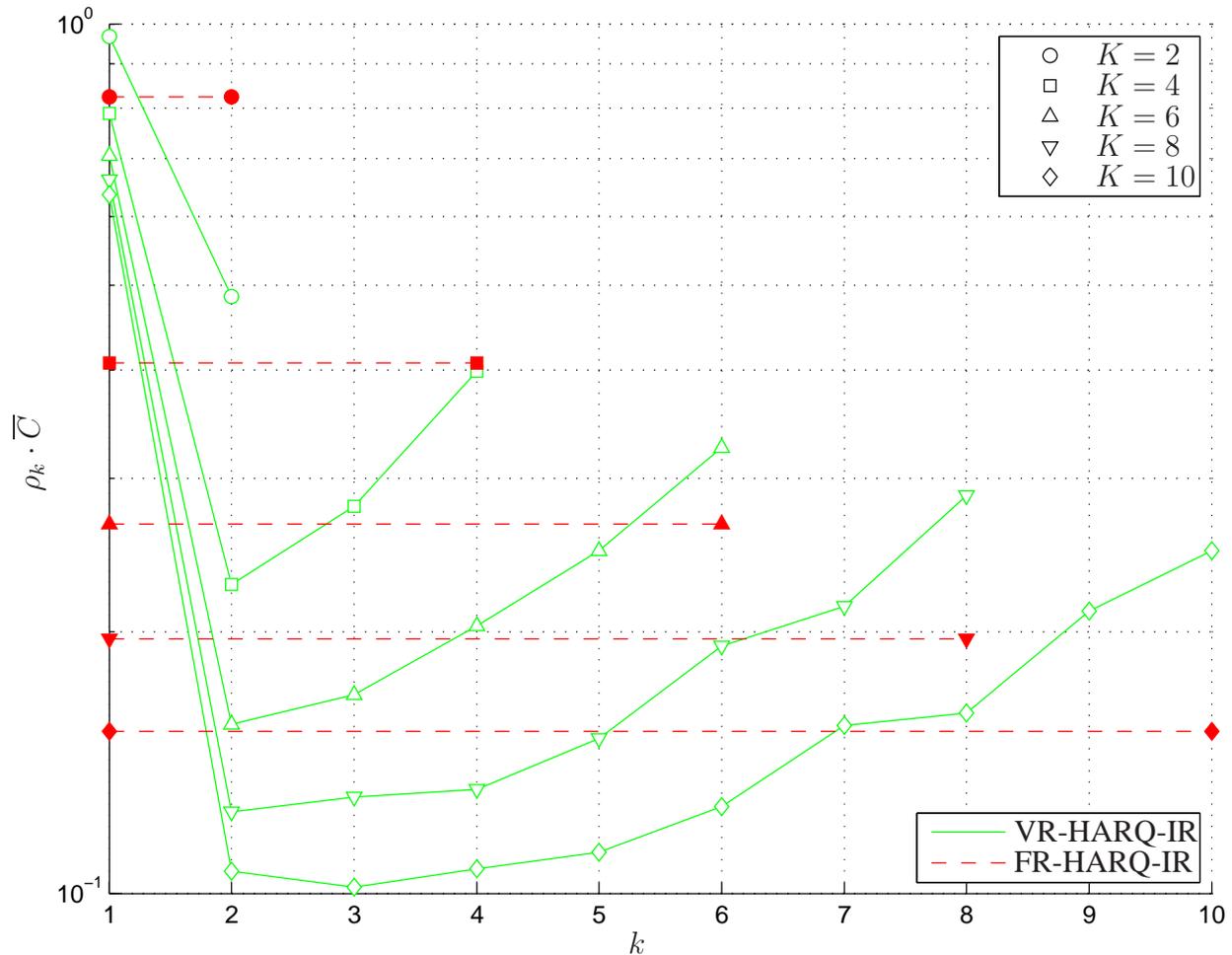}}
\caption{Optimal redundancy $\rho'_{k}=\rho\cd\ov{C}, k=1,\ld,K$ for VR-HARQ-IR (solid, green line) and FR-HARQ-IR (dashed, red line); $m=1$, $\ov{\gamma}=$10dB.}\label{Fig:RhoK}
\end{center}
\end{figure}

In Fig.~\ref{Fig:rates} we show the normalized redundancy $\rho'_{k}=\rho_{k}\cd\ov{C}$ directly proportional to the subcodewords' lengths $N_{\tr{s},k}$ (inversely proportional to the transmission rates $R_{k}$). We observe that the first transmission attempt of VR-HARQ-IR is carried out with the rate $R_{1}=1/\rho_{1}$ close to $\ov{C}$, while the rates of subsequent transmissions increase (i.e., the subcodewords are shorter) and decrease again for $k$ approaching $K$. This relationship holds for all $\ov{\gamma}$ and $m$ and may be observed in the IR and CHASE schemes. In Fig.~\ref{Fig:RhoK}, we reproduce similar results for VR-HARQ-IR and  FR-HARQ-IR  but for different values of $K$. The same ``profile''  of the redundancy is obtained for all $K$ and we may also appreciate that the values of $\rho'_{k}$ are decreasing with $K$, which is consistent with the optimal behaviour for $K\rightarrow\infty$, when the optimal sequence of $\rho_{k}$ should be non-increasing \cite[Lemma~1]{Visotsky03}. For the FR-HARQ-IR, we observe that $\rho'_{1}$ decreases with $K$. Recall that, according to the proof in \cite[Appendix~C]{Caire01}, when $K\rightarrow\infty$ the throughput-maximizing redundancy $\rho'_{k}=\rho_{1}\ov{C}$  should tend to $\frac{1}{K}$. 

\begin{figure}[tb]
\psfrag{xlabel}{$k$}
\psfrag{ylabel}{$f_{k}$}
\psfrag{XXXXX1}{$K=2$}
\psfrag{XXXXX2}{$K=4$}
\psfrag{XXXXX3}{$K=6$}
\psfrag{XXXXX4}{$K=8$}
\psfrag{XXXXX5}{$K=10$}
\psfrag{YYYYYYYYYY1}{VR-HARQ-IR}
\psfrag{YYYYYYYYYY2}{FR-HARQ-IR}
\begin{center}
\scalebox{1}{\includegraphics[width=1\linewidth]{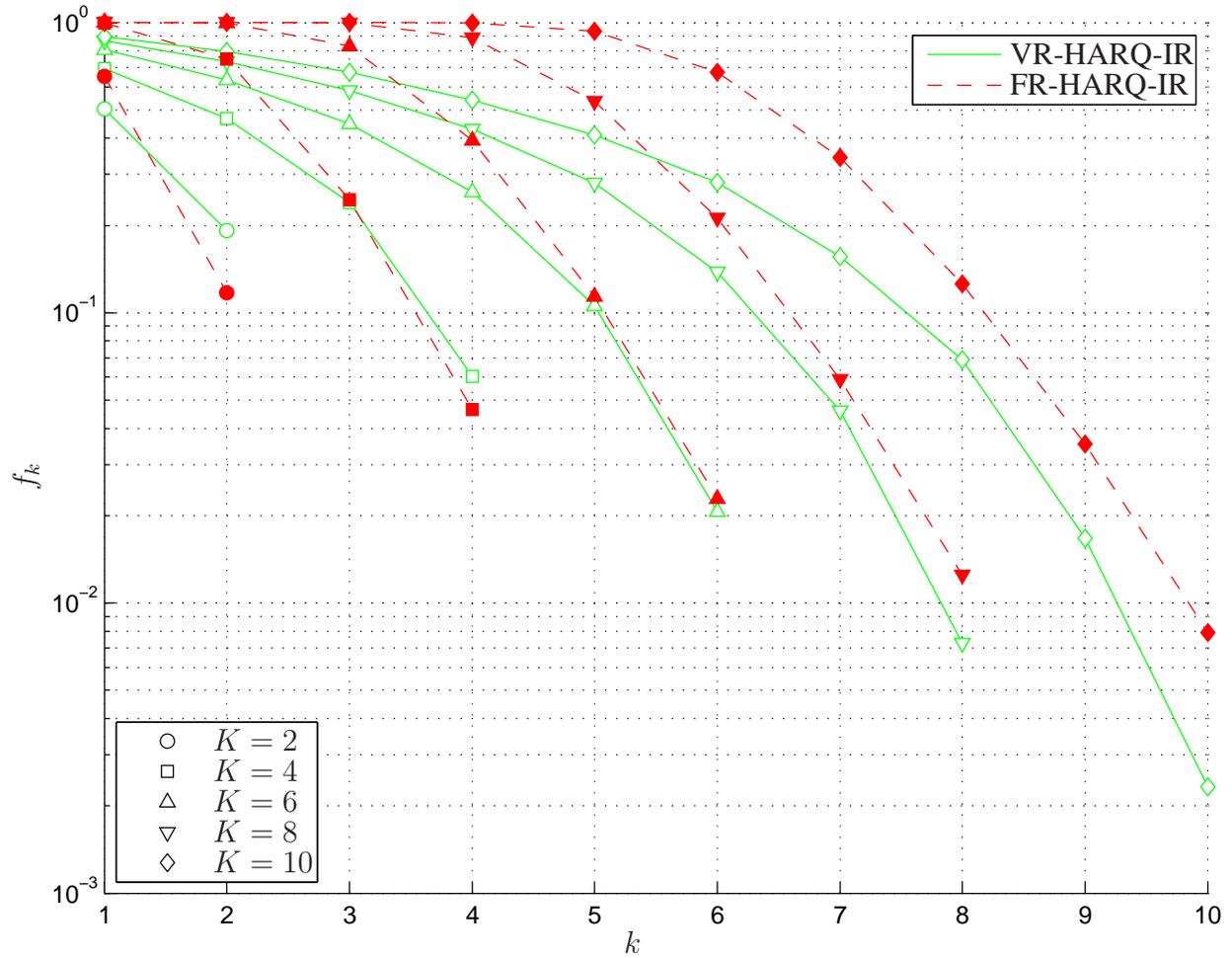}}
\caption{Outage values $f_{k}, k=1,\ld,K$ for VR-HARQ-IR (solid, green line) and FR-HARQ-IR (dashed, red line); $m=1$, $\ov{\gamma}=$10dB.}\label{Fig:fK}
\end{center}
\end{figure}

The decreasing-increasing behaviour of the values $\rho'_{k}$ can be interpreted from \eqref{TH.final}  combining the results of Fig.~\ref{Fig:RhoK} with those in Fig.~\ref{Fig:fK} showing the values of the decoding failure $f_{k}, k=1,\ld, K$. Namely, as we strive to make $\eta_{K}$ approach closely $\ov{C}$, from \eqref{TH.final} we conclude that redundancy/rate should be allocated so that $\ov{C}\cd(\rho_{1}+\sum_{k=2}\rho_{k}\cd f_{k-1})=\rho'_{1}+\sum_{k=2}\rho'_{k}\cd f_{k-1}$ grown to be as close as possible to $1-f_{K}\approx 1$. Immediately we conclude that we have to use  $\rho'_{1}<1$ (transmission rate $R_{1}>\ov{C}$) but the behaviour of optimal values $\rho'_{k}, k>1$ depends on how the values $f_{k}$ evolve with $\rho'_{k}$. 

In the particular case of $K\rightarrow\infty$, as long as the receiver ``accumulates'' the redundancy, the optimal values $\rho'_{k}$ should be decreasing with $k$ \cite{Visotsky03}\footnote{Remember that for HARQ-I, i.e., when the receiver discards the redundancy of past transmission attempts, the optimal solution is $\rho_{k}\equiv\rho_{1}$}. Thus, the fact that $\rho'_{k}$ increases with $k$ (here: for $k>2$) is due to the truncation (finite $K$) and reflects the fact that not only the denominator of \eqref{TH.final} should be minimized but also we have to guarantee that the value of $f_{K}$ remains small. 

Also, since $f_{k}$ decreases mach faster in HARQ-IR than it does in HARQ-CHASE (due to lack of additional information coded symbols conveyed in the subsequent transmission attempts), the optimal values of $\rho'_{k}, k>1$ can be smaller for VR-HARQ-IR than they are for VR-HARQ-CHASE so, as shown in Fig.~\ref{Fig:rates} the variation of the redundancy is less pronounced.

In Fig.~\ref{Fig:fK} we can also observe that for sufficiently large $K$ ($K\geq 4$), the probability of outage $f_{\IR,K}$ in VR-HARQ-IR is smaller than in the case of FR-HARQ-IR. For other values of $m$ and $\gamma$ the same property was consistently observed which is another clear advantage of VR-HARQ-IR over FR-HARQ-IR.

\begin{figure}[tb]
\psfrag{xlabel}{$\ov{\gamma}$}
\psfrag{ylabel}{$K_\tr{avg}$}
\psfrag{XXXXX1}{$K=2$}
\psfrag{XXXXX2}{$K=4$}
\psfrag{XXXXX3}{$K=8$}
\psfrag{YYYYYYYYYY1}{VR-HARQ-IR}
\psfrag{YYYYYYYYYY2}{FR-HARQ-IR}
\begin{center}
\scalebox{1}{\includegraphics[width=1\linewidth]{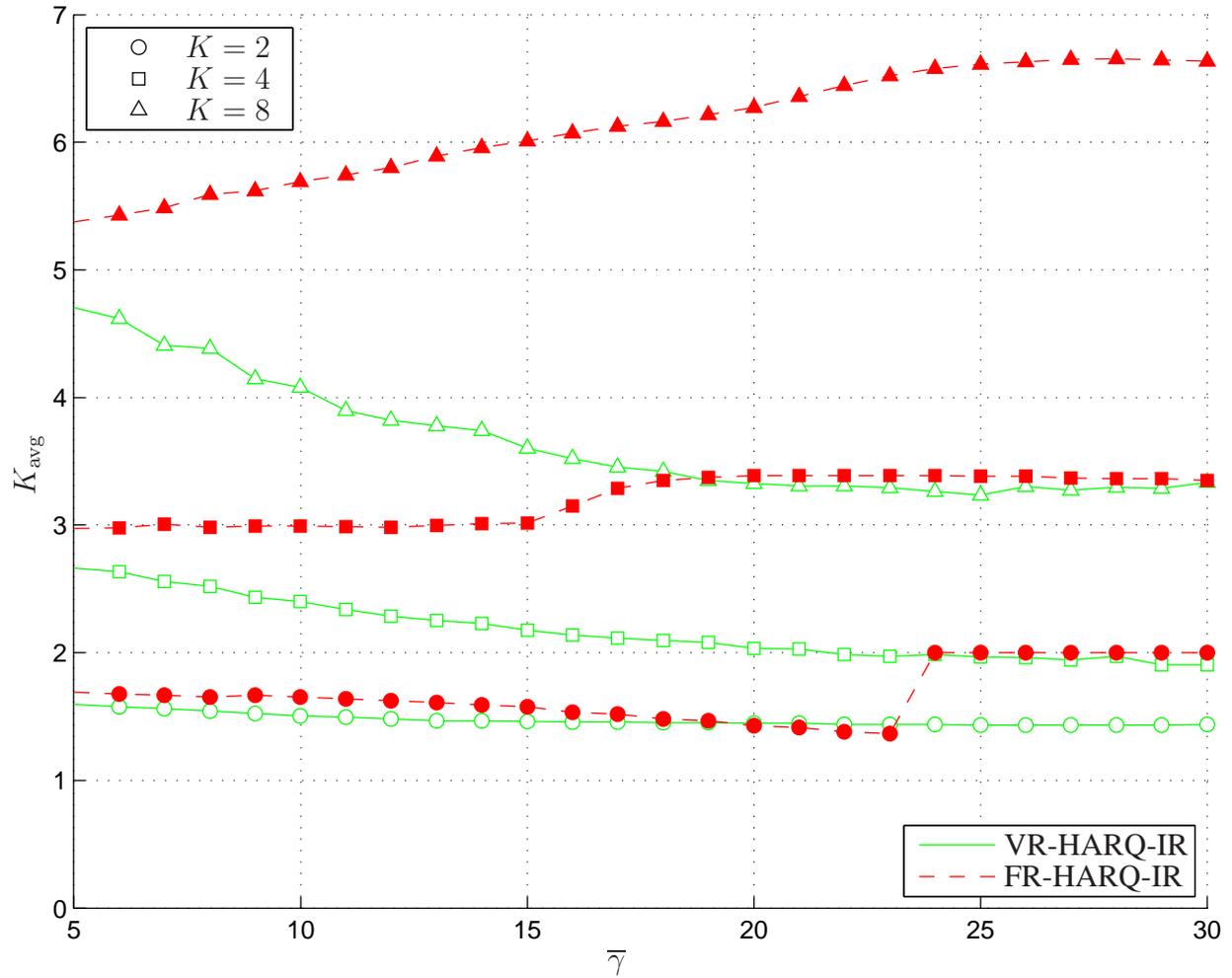}}
\caption{Average number of transmissions $K_\tr{avg}$ for VR-HARQ-IR (solid, green line) and FR-HARQ-IR (dashed, red line); $m=1$.}\label{Fig:Kavg}
\end{center}
\end{figure}

Another consequence of using short sub-codewords for all transmission attempts in FR-HARQ-IR is that the mutual information accumulates ``slowly'' with the retransmissions. Consequently, the failures in the initial transmissions occur more likely than in the VR-HARQ-IR, where the first transmission is done with the rate $R_{1}$ close to $\ov{C}$. This impacts the average number of transmissions which we calculate as
\begin{align}\label{Navg} 
  K_\tr{avg}=1+\sum_{k=1}^{K-1}f_{k}
\end{align}
and show in Fig.~\ref{Fig:Kavg}. 

We can appreciate that when the number of transmission $K$ grows, the average number of transmissions $K_\tr{avg}$ increases as well but is significantly greater for fixed-rate HARQ-IR: it practically doubles for $K=8$ and $\gamma=30$dB. Since the average number of transmissions is related to the packet delivery delay (as retransmission can be done only in separate frames), VR-HARQ-IR --besides the increased throughput-- offers an additional advantage  over FR-HARQ-IR.

\section{Conclusions}\label{Sec:Conclusions}
In this paper we have analyzed HARQ with incremental redundancy (HARQ-IR) for transmissions over block-fading channels. We have proposed an efficient method to allocate the optimal ratesand have demonstrated that the variable-rate HARQ-IR  provides gains over the fixed-rate HARQ-IR in terms of increased throughput, lower outage, and decreased average number of transmissions.
 
\section*{Acknowledgment}
The authors thank Dr. M. Benjillali (INPT, Rabat, Morocco) for his critical reading and Prof. J. Benesty (INRS-EMT, Montreal) for the suggestions leading to the simplification of the outage calculation in Sec.~\ref{Sec:IR}.


\begin{thebibliography}{10}
\providecommand{\url}[1]{#1}
\csname url@samestyle\endcsname
\providecommand{\newblock}{\relax}
\providecommand{\bibinfo}[2]{#2}
\providecommand{\BIBentrySTDinterwordspacing}{\spaceskip=0pt\relax}
\providecommand{\BIBentryALTinterwordstretchfactor}{4}
\providecommand{\BIBentryALTinterwordspacing}{\spaceskip=\fontdimen2\font plus
\BIBentryALTinterwordstretchfactor\fontdimen3\font minus
  \fontdimen4\font\relax}
\providecommand{\BIBforeignlanguage}[2]{{%
\expandafter\ifx\csname l@#1\endcsname\relax
\typeout{** WARNING: IEEEtran.bst: No hyphenation pattern has been}%
\typeout{** loaded for the language `#1'. Using the pattern for}%
\typeout{** the default language instead.}%
\else
\language=\csname l@#1\endcsname
\fi
#2}}
\providecommand{\BIBdecl}{\relax}
\BIBdecl

\bibitem{Brayer68}
K.~Brayer, ``Error control techniques using binary symbol burst codes,''
  \emph{{IEEE} Trans. Commun.}, vol.~16, no.~2, pp. 199 --214, Apr. 1968.

\bibitem{Caire01}
G.~Caire and D.~Tuninetti, ``The throughput of hybrid-{ARQ} protocols for the
  {G}aussian collision channel,'' \emph{{IEEE} Trans. Inf. Theory}, vol.~47,
  no.~5, pp. 1971--1988, Jul. 2001.

\bibitem{Wu10}
P.~Wu and N.~Jindal, ``Performance of hybrid-{ARQ} in block-fading channels: A
  fixed outage probability analysis,'' \emph{{IEEE} Trans. Commun.}, vol.~58,
  no.~4, pp. 1129 --1141, Apr. 2010.

\bibitem{Tuninetti11}
D.~Tuninetti, ``On the benefits of partial channel state information for
  repetition protocols in block fading channels,'' \emph{CoRR}, vol.
  abs/1102.4085, 2011.

\bibitem{Visotsky03}
E.~Visotsky, V.~Tripathi, and M.~Honig, ``Optimum {ARQ} design: a dynamic
  programming approach,'' in \emph{Proc. IEEE International Symposium on
  Information Theory}, Jun. 2003, p. 451.

\bibitem{Uhlemann03}
E.~Uhlemann, L.~K. Rasmussen, A.~Grant, and P.-A. Wiberg, ``Optimal
  incremental-redundancy strategy for type-{II} hybrid {ARQ},'' in \emph{Proc.
  IEEE International Symposium on Information Theory}, 2003, p. 448.

\bibitem{Cheng03}
J.-F. Cheng, Y.-P. Wang, and S.~Parkvall, ``Adaptive incremental redundancy,''
  in \emph{{IEEE} Veh. Tech. Conf.}, Orlando, Florida, USA, Oct. 2003, pp.
  737--741.

\bibitem{LiuR03}
R.~Liu, P.~Spasojevic, and E.~Soljanin, ``On the role of puncturing in hybrid
  {ARQ} schemes,'' in \emph{Proc. IEEE International Symposium on Information
  Theory}, Jun. 2003, p. 449.

\bibitem{Visotsky05}
E.~Visotsky, Y.~Sun, V.~Tripathi, M.~Honig, and R.~Peterson,
  ``Reliability-based incremental redundancy with convolutional codes,''
  \emph{{IEEE} Trans. Commun.}, vol.~53, no.~6, pp. 987 -- 997, Jun. 2005.

\bibitem{Gopalakrishnan08b}
N.~Gopalakrishnan and S.~Gelfand, ``Rate selection algorithms for {IR} hybrid
  {ARQ},'' in \emph{2008 {IEEE} Sarnoff Symposium}, Princeton, NJ, USA, Apr.
  2008, pp. 1--6.

\bibitem{Kim11}
S.~M. Kim, W.~Choi, T.~W. Ban, and D.~K. Sung, ``Optimal rate adaptation for
  hybrid {ARQ} in time-correlated {R}ayleigh fading channels,'' \emph{{IEEE}
  Trans. Wireless Commun.}, vol.~10, no.~3, pp. 968 --979, Mar. 2011.

\bibitem{Shen09}
C.~Shen, T.~Liu, and M.~Fitz, ``On the average rate performance of hybrid-{ARQ}
  in quasi-static fading channels,'' \emph{{IEEE} Trans. Commun.}, vol.~57,
  no.~11, pp. 3339 --3352, Nov. 2009.

\bibitem{Tuninetti07}
D.~Tuninetti, ``Transmitter channel state information and repetition protocols
  in block fading channels,'' in \emph{{IEEE} Information Theory Workshop, ITW
  '07}, California, USA, Sep. 2007, pp. 505--510.

\bibitem{Gamal06}
H.~Gamal, G.~Caire, and M.~Damen, ``The {MIMO} {ARQ} channel:
  Diversity--multiplexing--delay tradeoff,'' \emph{{IEEE} Trans. Inf. Theory},
  vol.~52, no.~8, pp. 3601 --3621, Aug. 2006.

\bibitem{Nguyen10}
K.~D. Nguyen, L.~K. Rasmussen, A.~G. i~Fabregas, and N.~Letzepis, ``{MIMO}
  {ARQ} with multi-bit feedback: Outage analysis,'' \emph{CoRR}, vol.
  abs/1006.1162v2, 2010.

\bibitem{Gopalakrishnan08}
N.~Gopalakrishnan and S.~Gelfand, ``Achievable rates for adaptive {IR} hybrid
  {ARQ},'' in \emph{2008 {IEEE} Sarnoff Symposium,}, Apr. 2008, pp. 1--6.

\bibitem{Kim07}
T.~Kim and M.~Skoglund, ``On the expected rate of slowly fading channels with
  quantized side information,'' \emph{{IEEE} Trans. Commun.}, vol.~55, no.~4,
  pp. 820--829, Apr. 2007.

\bibitem{Goldsmith97_b}
A.~J. Goldsmith and P.~Varaiya, ``Capacity of fading channels with channel side
  information,'' \emph{{IEEE} Trans. Inf. Theory}, vol.~43, no.~6, pp.
  1986--1992, 1997.

\bibitem{Liu04}
Q.~Liu, S.~Zhou, and G.~B. Giannakis, ``Cross-layer combining of adaptive
  modulation and coding with truncated {ARQ} over wireless links,''
  \emph{{IEEE} Trans. Wireless Commun.}, vol.~3, no.~5, pp. 1746--1755, Sep.
  2004.

\bibitem{Wang07}
X.~Wang, Q.~Liu, and G.~Giannakis, ``Analyzing and optimizing adaptive
  modulation coding jointly with {ARQ} for {QoS}-guaranteed traffic,''
  \emph{{IEEE} Trans. Veh. Technol.}, vol.~56, no.~2, pp. 710--720, Mar. 2007.

\bibitem{Zorzi96}
M.~Zorzi and R.~Rao, ``On the use of renewal theory in the analysis of {ARQ}
  protocols,'' \emph{{IEEE} Trans. Commun.}, vol.~44, no.~9, pp. 1077--1081,
  Sep 1996.

\bibitem{Malkamaki99}
E.~Malkamaki and H.~Leib, ``Coded diversity on block-fading channels,''
  \emph{{IEEE} Trans. Inf. Theory}, vol.~45, no.~2, pp. 771--781, Feb. 1999.

\bibitem{Cheng06}
J.~Cheng, ``Coding performance of hybrid {ARQ} schemes,'' \emph{{IEEE} Trans.
  Commun.}, vol.~54, no.~6, pp. 1017--1029, Jun. 2006.

\bibitem{Szczecinski10}
L.~Szczecinski, C.~Correa, and L.~Ahumada, ``Variable-rate transmission for
  incremental redundancy hybrid {ARQ},'' in \emph{IEEE Global
  Telecommunications Conference, GLOBECOM 2010}, Dec. 2010.

\bibitem{Bertsekas05_book}
D.~P. Bertsekas, \emph{Dynamic Programming and Optimal Control}, 3rd~ed.\hskip
  1em plus 0.5em minus 0.4em\relax Athena Scientific, 2005, vol.~1.

\end{thebibliography}

\end{document}